\documentclass[10pt]{article}

\usepackage{amsmath}
\usepackage{amssymb}

\usepackage{graphicx}

\usepackage{cite}

\usepackage{color}

\usepackage[labelfont=bf,labelsep=period,justification=raggedright]{caption}

\makeatletter
\renewcommand{\@biblabel}[1]{\quad#1.}
\makeatother

\date{}

\begin{document}

% Title must be 150 characters or less
\begin{flushleft}
{\Large
\textbf{Wikipedia information flow analysis reveals the scale-free architecture of the Semantic Space}
}
% Insert Author names, affiliations and corresponding author email.
\\
Adolfo Paolo Masucci$^{1,\ast}$,
Alkiviadis Kalampokis$^{2}$,
Victor Mart\'inez Egu\'iluz$^{1}$,
Emilio Hern\'andez-Garc\'ia$^{1}$
\\
\bf{1} Instituto de F\'isica Interdisciplinar y Sistemas Complejos,  Consejo Superior de Investigaciones Científicas - Universitat de les Illes Balears, Palma de Mallorca, Spain
\\
\bf{2} Department of Marine Sciences, University of the Aegean,  Mytilene, Lesvos, Greece
%\\
%\bf{3} Author3 Dept/Program/Center, Institution Name, City, State, Country
\\
$\ast$ E-mail: paolo@ifisc.uib.es
\end{flushleft}

% Please keep the abstract between 250 and 300 words
\section*{Abstract}
In this paper we extract the topology of the semantic space in its encyclopedic acception, measuring  the semantic flow between the different
entries of the largest modern encyclopedia, Wikipedia, and thus creating a directed complex network of semantic flows.
 Notably at the percolation threshold the semantic space is characterised by scale-free behaviour at different levels of complexity and this relates the semantic space to a wide range of  biological, social and linguistics phenomena. In particular we find that the cluster size distribution, representing the size of different semantic areas, is scale-free.   Moreover the  topology of the resulting semantic space is scale-free in the connectivity distribution and displays small-world properties. However its statistical properties do not allow a classical interpretation via a generative model based on a simple multiplicative process. After giving a detailed description and interpretation of the topological properties of the semantic space, we introduce a stochastic model of content-based network, based on a copy and mutation algorithm and on the Heaps' law, that is able to capture the main statistical properties of the analysed semantic space, including the Zipf's law for the word frequency distribution.
% Please keep the Author Summary between 150 and 200 words
% Use first person. PLoS ONE authors please skip this step.
% Author Summary not valid for PLoS ONE submissions.
%\section*{Author Summary}

\section*{Introduction}
The meaning of a word can be defined as an indefinite set of {\it interpretants}, which are other words that circumscribe the semantic content of the word they represent \cite{47}. In the same way each interpretant has a set of interpretants representing it and so on. Hence the indefinite chain of meaning assumes a {\it rhizomatic} shape  that can be represented and analysed via the modern techniques of network theory \cite{40}.

The semantic or conceptual space (SS hereafter) has already
been investigated by different approaches. A common
understanding within these approaches is that the SS is made up of
words or concepts that are connected by certain relationships.
Depending on the nature of these relationships different semantic
webs have already been  considered.
In the psycholinguistics approach the SS is often extracted via \emph{free word association}
game and a network  is constructed where two words are connected if they appear to be consecutive in a free word
association experiment \cite{38,Borge10}.
Other semantic webs are generated through linguistics approaches \cite{37}.
Among  others an interesting one is  based on the dictionary, where the relationships between words are set to be of synonymy, antonymy, belonging to the same category or class, etc. \cite{36,s1}. In all the
mentioned cases a scale-free topology and small-world
properties for the SS are found, suggesting an
intrinsic self-organising nature of the SS \cite{36}.
However it has been argued that networks derived by dictionaries  and representing the so called \emph{dictionary semantics}, characterised by scale-free distribution for the connectivity with exponents smaller than -2,  reflect the properties of language use  more than the properties of the SS \cite{52,50}.

In contrast to the dictionary representation of the  SS, it has been suggested that the meaning of a sign, where a sign can be a word, a concept, etc. can be recovered within an \emph{encyclopedic model}, where every sign is specified by a set of other signs that interpret it \cite{22}.
``This notion of interpretants is fertile because it shows how semiotic processes, via continuous movements that refer a sign to other signs or sign chains, circumscribe the meanings in an asymptotic way. They never touch them, and make them accessible via other cultural units [...]. In this way an open system of connections between different signs is created that takes the shape of a rhizome \cite{23}" \cite{22}.

Hence, in its \emph{encyclopedic semantics} acception, the SS can be interpreted as a metapopulation system where each page of an encyclopedia is a population of interpretants/words characterising some meanings. Then the structure of SS assumes a dynamical connotation, typical of population dynamics, where the different concepts are born and grow in time, exchanging  and inheriting attributes from other concepts\footnote[1]{It is interesting to notice how Deleuze and Guattari foresaw the very essence of the semantic machine not as a machine producing meaning, but as a machine producing its own structure \cite{dg}.}.

In this work we attempt to extract the SS in its encyclopedic semantics acception. Following the semiotics rationale described above, we consider each page of an encyclopedia as a population of interpretants and we  measure the correlations between each pair of pages of that encyclopedia in terms of \emph{directional semantic flows}. In particular we  analyse a whole dump of Wikipedia. Wikipedia is not only the largest encyclopedia existing nowadays, but  it is an \emph{open encyclopedia} with its pages always growing in size and number, thus it  represents well the idea of encyclopedic semantics expressed above. The resulting network is  a directed network of semantic flows between the different concepts that are present in an encyclopedia and thus portrays a snapshot of the dynamics of meaning in that representation of the SS.

The concept of \emph{information flow}, as it is used in this context, is  introduced in \cite{s2} to indicate the correlations between populations whose elements are defined by abstract attributes. Those populations can be social, biological or, as in this case, made of words. We choose the use of the term ``information flow", instead of distance between probability distributions or correlations, because in those systems correlations are often caused by migration or inheritance of a part of a population to another one. Thus the very term of information flow, that can be ethnical, genetical or, as in this case, semantic, gives an idea of such a dynamical process, where to a movement of elements from a population to another one, it corresponds to a flow of information in the attribute space where those elements are defined.

 \subsection{The dataset} We  consider the articles of a complete snapshot of English Wikipedia dated  June 2008 \cite{wp}, consisting of $N \approx 2 \cdot10^6$ entries. To process our dataset first of all we get rid of  redirection pages. Then, in order to analyse the semantic content of the encyclopedia, we process the text,  cleaning it of punctuation and of the so called \emph{structural words} like articles, pronouns, common adverbs, etc. \cite{26,27}. In fact those words are very frequent in each page and often don't contribute to its semantic characterisation. After  that we \emph{lemmatise} the text, transforming the different words in their singular form or in their infinite form if they are verbs \cite{39}.
The resulting set of lemmas defines the interpretants or \emph{attribute} space where the different pages are defined and each Wikipedia page comes out to be defined by its lemmas frequency distribution  and by its size.

 In order to compute the directional semantic flow between the pages we use the method introduced in \cite{s2}. This method is very general and allows the extraction of a directed information flow network from a set of populations whose elements are defined by an $n$-dimensional symbolic attribute vector. It is  based on the Jensen-Shannon divergence \cite{1} and within an information theory approach it is able to measure the amount of information flow within a set of  populations of different sizes,  defined in a symbolic attribute space. Moreover, using concepts derived from geographical segregation, the methodology in \cite{s2} is able to infer the directionality of the information flow. More details are given in the Materials and Methods section.

The resulting network representing the SS, as we show below,      displays scale invariant structures and small world properties, revealing a hierarchical SS, where the semantic clusters are strongly connected and communication between different areas of knowledge is fast.

%\section*{Figure Legends}
%%%%%%%%%%%%%%%%%%%%%%%%%%%%%%%%%%%%%%%%%%%%%%%%%%%%%%%%%%%%%%%% fig.1
\begin{figure}[!ht]\center
               \includegraphics[width=0.6\textwidth]{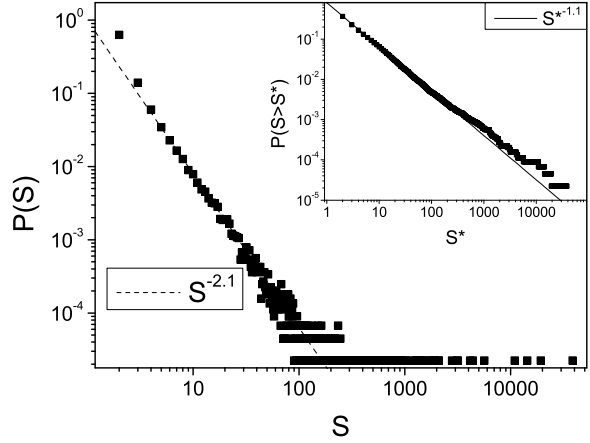}
 \caption{\label{f1} \textbf{Cluster size distribution of the semantic space.} Cluster size distribution $P(S)$ of the semantic
 network at the percolation threshold.  In the inset we show the cumulative distribution $P(S>S^*)$.}
 \end{figure}
 
% Results and Discussion can be combined.
\section*{Results}
\subsection*{Topology of the Semantic Space}

To build the network the directional semantic flow is measured  between all the entry pairs. Then  the entry pairs are ordered by the increasing values of their semantic distance, and  a network of entries is defined considering two pages as linked when their semantic distance is smaller than a given threshold.

By increasing the value of the threshold we obtain a growing network where the first links to form are the strongest in a semantic sense. As the threshold is  increased further, very well connected clusters form, each cluster representing different semantic areas. A significative threshold to analyse the network representing the SS is the \emph{percolation threshold} (PT hereafter), when a giant cluster forms and a phase transition
happens \cite{30,40}.

%%%%%%%%%%%%%%%%%%%%%%%%%%%%%%%%%%%%%%%%%%%%%%%%%%%%%%%%%%%%%%%%%%%%%%%%%%Fig. 2
\begin{figure*}[!ht]\center
              \includegraphics[width=0.9\textwidth]{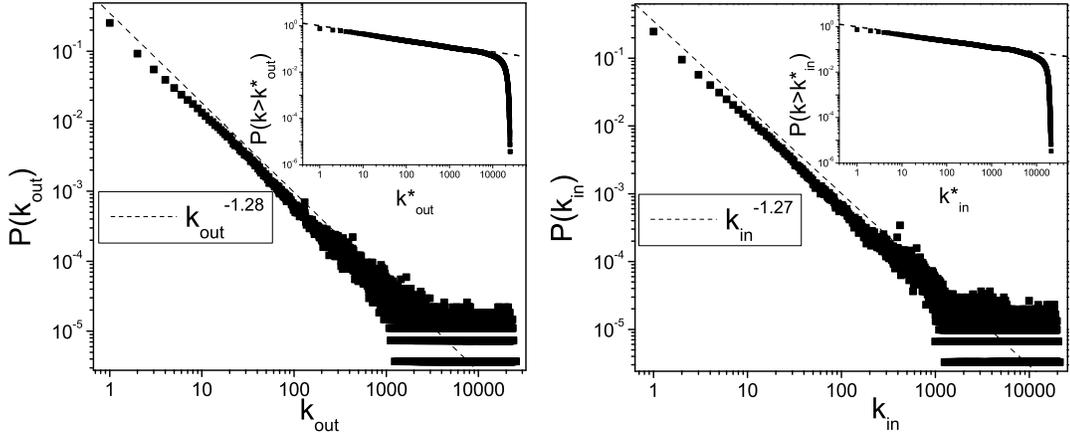}
 \caption{\label{f2} \textbf{Connectivity distribution of the semantic space.} Out-degree distribution $P(k_{out})$ (left panel) and in-degree distribution $P(k_{in})$ (right panel) of the semantic network at the percolation threshold.  In the insets the corresponding cumulative degree distributions $P(k>k^*)$ are displayed.}
 \end{figure*}

 The SS network reaches its PT  at  approximately 362000 pages, when the two main clusters merge to form a giant cluster of 57800 pages. At the PT the network has $L\approx 3\cdot 10^8 $ links with an average degree $\langle k \rangle \approx 1743$. The very large average degree means that the clusters are very densely connected.   The network is composed by 44500 disconnected clusters showing scale invariant cluster size distribution, $P(S)\propto S^{-2.1}$, with a fat tail (Fig.\ref{f1}).

  The scale-free cluster size distribution is the first important property we find for the SS.
   It has been shown that in a random growing network at the PT the cluster size distribution decays faster than a power law,  $P(S)\propto 1/( S^3\ln^2 S)$ \cite{40,kk}. Then it can be argued that the scale-free behaviour we find for the cluster size distribution is not an effect of a random growing network at  percolation, but a peculiar property of the SS.

   As a matter of fact at this threshold almost each cluster represents a well defined semantic area. Hence the scale-free distribution implies a hierarchy between the semantic areas and gives us a picture of the structure of the SS.

   The  largest clusters, representing the greater body of the SS, are composed of large taxonomies, such as geographical places, biological species, etc...
   The largest cluster is made of 38500 pages and consists of geographical places of USA, such as villages, cities, rivers, etc... The second largest cluster is made of 19300 pages and consists of taxonomies of living species as animal, plants, insects, bacteria, etc...The third largest cluster is mainly made of Romanian geographical entries, the fourth by French cities and villages and so on.
   In each of these clusters the pages are very simple and have a structure very similar to each other. A typical example of these kinds of pages is the \emph{Canarium Zeylanicum} page, that is in the second largest cluster: ``Canarium Zeylanicum is a species of flowering plant in the frankincense family, Burseraceae, that is endemic to Sri Lanka.". The content word lemmas of this page are: ``Canarium Zeylanicum  specie flower plant  frankincense family Burseraceae endemic Sri Lanka". This page easily connects with all pages containing ``specie flower plant endemic Sri Lanka", hence forming a taxonomy with other pages as the \emph{Mastixia Nimali} page: ``Mastixia Nimali is a species of plant in the Cornaceae family. It is endemic to Sri Lanka.".
It is interesting to notice how the passage from a taxonomic page to another resides in the mutation of a few rare words.

   Then  clusters are found at all scales of magnitude, consisting of any semantic area one can possibly think of and generally the greater is the complexity of the page, the smaller is the cluster  which it belongs to. There are clusters of football players,  small clusters of different kind of bicycles, ethnicity clusters,  language family clusters, singers, technology, religions, etc...

The out-degree and in-degree distribution of the network at the PT are scale-free  with a very slow decay, characterised by exponents: $\gamma_{out}\approx -1.28$ and $\gamma_{in}\approx-1.27$ (see Fig.\ref{f2}). The distributions are scale invariant until very large scales where a sharp cut-off appears, revealing that the SS is characterised by structures at all the scales. The giant component of the network has a directed diameter $d=20$ that is of the order of the logarithm of the cluster size. Moreover its average clustering coefficient is $\langle C \rangle=0.87$, that is larger than the clustering coefficient of a random network of the same size, $\langle C\rangle_{RAND}=0.17$, revealing local \emph{small-world} properties of the SS \cite{43}.

%%%%%%%%%%%%%%%%%%%%%%%%%%%%%%%%%%%%%%%%%%%%%%%%%%%%%%%%%%%%%%%%%%%%%%%%%%Fig. 3
\begin{figure}[!ht]\center
             \includegraphics[width=0.6\textwidth]{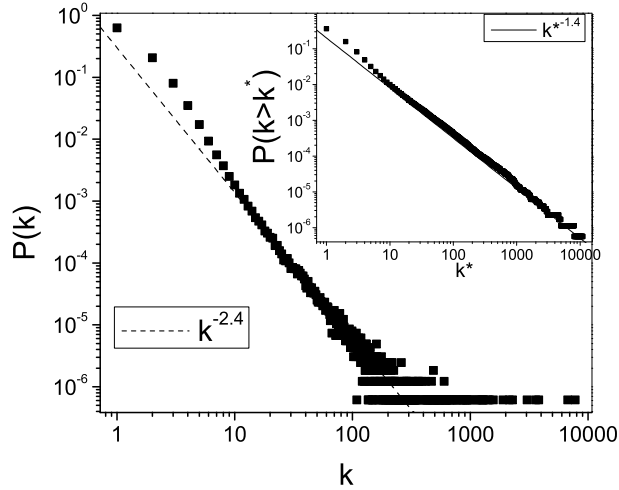}
 \caption{\label{f3} \textbf{Connectivity distribution of the minimum spanning tree of the semantic space.} Degree distribution $P(k)$ for the undirected minimum spanning tree for the whole network representing the semantic space.
  In the insets the cumulative degree distribution $P(k>k^*)$ is displayed.}
 \end{figure}

If the description at the cluster level  represents the main body of the SS,  the \emph{minimum spanning tree} (hereafter MST) represents its backbone.
The MST  of a weighted network is an acyclic graph that has all the vertices of the network and that minimises the sum of the distances between the pages \cite{41}. It represents the skeleton of the network and in a sense it represents how semantic information best flows throughout the SS.

\begin{figure*}[!ht]\center
                    \includegraphics[width=0.9\textwidth]{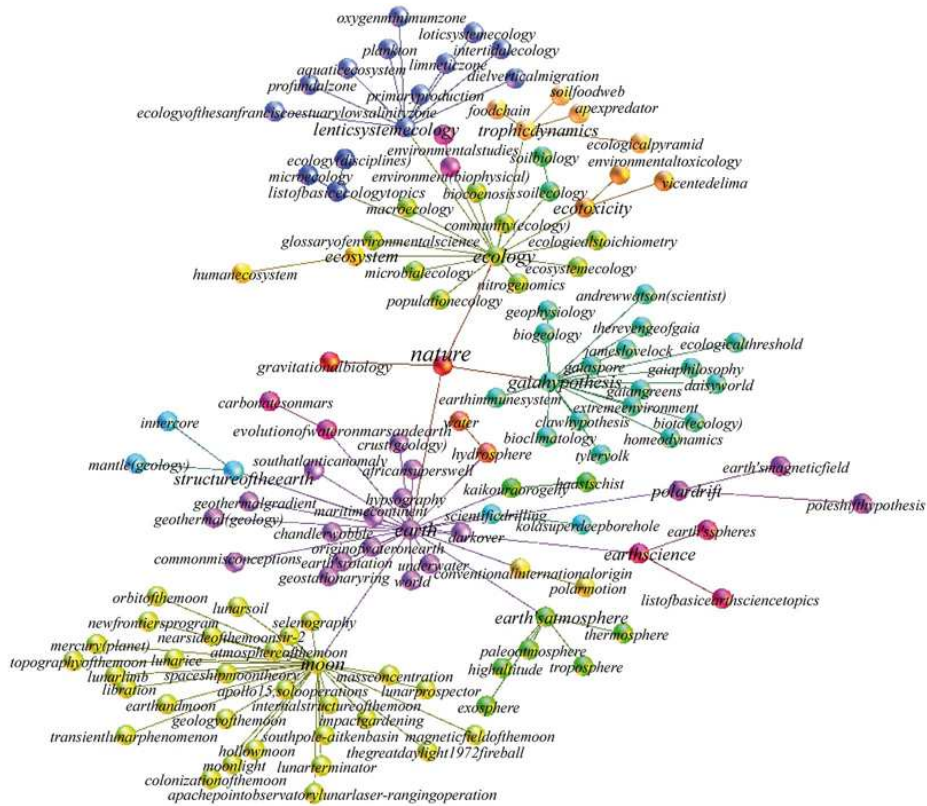}
 \caption{\label{f4} \textbf{A portion of the minimum spanning tree of the semantic space.} A portion of the undirected minimum spanning tree of the network representing  the semantic space in the neighbourhood of the entry \emph{nature} until its  third neighbour. The nodes represent different Wikipedia entries, while the edges  represent a semantic flow between the different entries.
 The color partition is based on the nodes modularity classes. Figure realised with the opensource software Gephi \cite{ge}.
 }
 \end{figure*}

We compute the undirected MST of the complete network of Wikipedia via the Prim's algorithm \cite{42}. The degree distribution of the MST is scale free with exponent   -2.4 and a fat tail (see Fig.\ref{f3}). Again the scale-free behaviour of the degree distribution tells us about the hierarchical structure of the MST of the SS. If we  glimpse at Fig.\ref{f4}, where  a small portion of the MST centred on the Wikipedia entry \emph{nature} is shown, we can have a rough idea of how this hierarchy organises itself. A very general concept, such as ``nature", hasn't got a lot of connections, but it is an important bridge for the semantic flow between less complex concepts. Those less complex concepts are in general more connected and eventually form taxonomies, which are hubs in the MST.

From what has been said we can draw a general picture of the SS, as a space whose body is mainly composed of simple concepts that are densely clustered in taxonomies or classifications. Then, at higher levels, more complex concepts form, creating smaller semantic clusters. This hierarchy goes further, in a scale-free fashion, until the more general and elaborated concepts emerge and those create an architecture of semantic flow channels that spans through the whole SS.

The values of the exponents of the degree distributions are too large to be explained by standard growing network models based on preferential attachment \cite{49}. For the character of the system and its statistical properties, the emerging topology of the SS is more likely to be represented  by a new class of models of stochastic \emph{content-based networks} of the type presented in \cite{48b,48,48c}, with the difference that in the case of Wikipedia the correlations generated by the zipfean distribution of content words \cite{27} play an important role on the topology of the system as it is explained below. This observation relates the topology of the SS to a wider range of biological phenomenology \cite{48c}.

%%%%%%%%%%%%%%%%%%%%%%%%%%%%%%%%%%%%%%%%%%%%%%%%%%%%%%%%%%%%%%%%%%%%%%%%%%Fig. 4

\subsection*{The model}

The complexity of the system we are considering is  large, since it relates the phenomenology of different topics page writing to the topology of the macroscopic system of the SS.

Here we present a descriptive model that is able to catch the properties of the SS at three different levels of complexity. In particular it is able to reproduce the scale-free cluster size distribution, the exponents for the out and in-degree distribution  and the Zipf's law for the word frequency distribution \cite{46}.
It is a growing stochastic model  of  content-based network generated by  a copy and mutation algorithm. This is intended on one hand to create a multiplicative process \emph{a la} Simon \cite{l3}, to reproduce the scale-free cluster distribution and on the other hand to create, via mutation, very well connected taxonomic clusters to reproduce the very low exponent of the degree distribution. In the overall process a Heaps'-like law \cite{l4} for the text growth is imposed to produce correlations between pages and to allow a phase transition and this finally generates the Zipf's law for the word frequency distribution
\footnote[2]{
 Written human language displays a fascinating puzzle of  empirical regularities. Among them, the Heaps' law states that the vocabulary $V$ of a written text is a function of its size $L$,
 $V(L)\propto L^\beta$,
%\begin{equation}
%V(L)\propto L^\beta,
%\end{equation}
with $0<\beta<1$.
 The Heaps' law is strictly related to the Zipf's law for the word frequency distribution.
 The Heaps' law has been analytically and algorithmically derived from the Zipf's law \cite{h1,h3}. In \cite{h2} there is a derivation of the Zipf's law from the Heaps' law, even if it is not straightforward.

In network theory a written text can be represented by a network whose vertices are the words and two vertices are linked if they are adjacent in the text \cite{h4,h5}.
A convenient way to model a growing text in network theory is to assume that at each time step $t$ a new word  and  a fraction of old words
$\alpha\cdot t$ is introduced in text, possibly preserving the \emph{eulerianity} of the system \cite{h6}.
Hence in this representation the discrete time $t$ represents the size of the vocabulary. At each time step $\alpha\cdot t+1$ words are
introduced in the text and the size $L$ of the text is a function of the vocabulary size: $L(t)=\frac{\alpha}{2}t^2+t$.
\begin{equation}
L(t)=\frac{\alpha}{2}t^2+t.
\end{equation}
}.

We give the details of the model in the Materials and Methods section, while here we just give its general description. Each page in the model is considered to be a collection of words. Those words can be new, extracted from a potentially infinite vocabulary, or old, picked up randomly from already written pages. The balance between new and old words is set to respect the Heaps'-like law between  text size and  vocabulary size. Each page is generated with an \emph{invariant part}, whose size is random, that is a fixed portion of the page that doesn't change when the page mutates.

The model starts with a few pages of random words. Then, at each time step, we generate a new page of  words, that, as explained above, are in part new words and in part  words picked up randomly from already written pages, keeping the balance between text size and vocabulary size via the Heaps'-like law. Then, at each time step, we generate $M$ new pages by copying $M$ old pages, keeping unchanged their invariant part and mutating their variant part with some old or new words, always considering the balance between text size and vocabulary size via the Heaps'-like law.

 The generation of new pages in the model is intended to mimic the appearance of new pages in the encyclopedia, that are formed partially by a new vocabulary and partially by words that are inherited by already existing pages. Beside, the generation of pages via the mutation mechanism allows to generate the different growing taxonomies and in this way to mimic the phenomenology observed in the real encyclopedia, where different pages belonging to the same taxonomy differ only by a few rare words.

\begin{figure}[!ht]\center
                \includegraphics[width=0.8\textwidth]{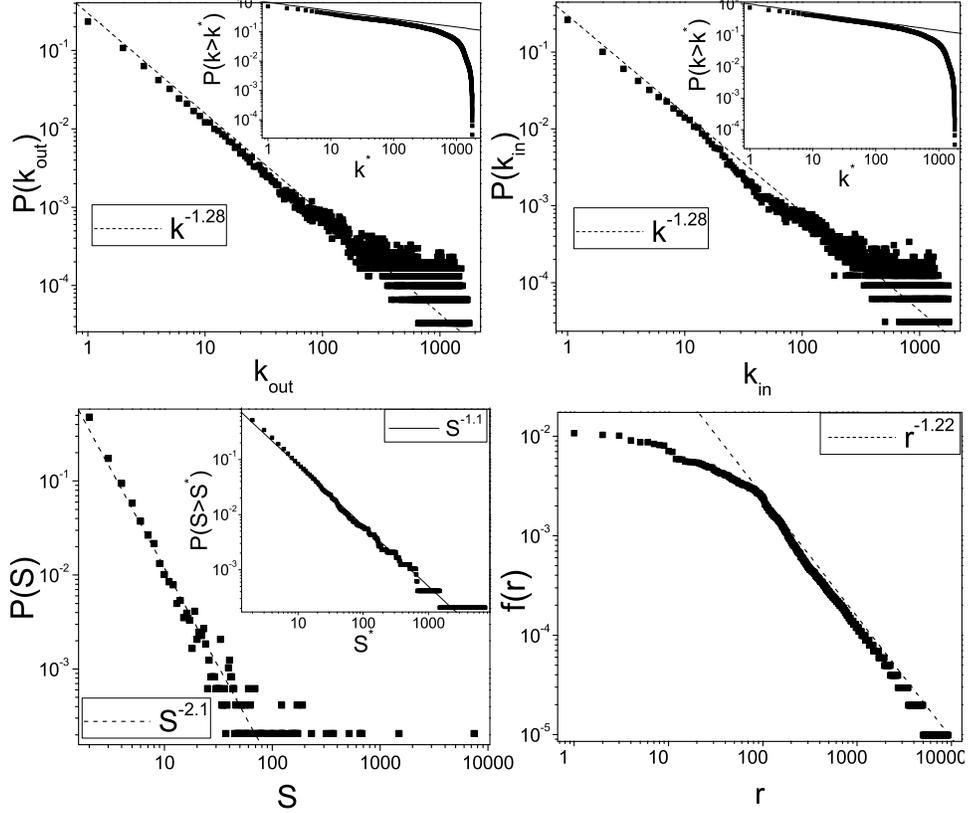}
 \caption{\label{f5}   \textbf{The stochastic model representing the semantic space.} Results for the simulation of the stochastic model representing the semantic space. This is a simulation of a \emph{toy-model} for an encyclopedia of $5\cdot 10^4$ pages with size $l$ log-normally distributed, with first moment $\overline l=20$ and second moment $l_\sigma=0.5$. The parameter of the model are $\alpha=0.001$, $p=0.7$ and $M=5$. In the top panels  the out-degree distribution $P(k_{out})$ (left panel) and the in-degree distribution $P(k_{in})$ (right panel) of the semantic network at the percolation threshold are shown. The corresponding cumulative distributions $P(k>k^*)$ are displayed in the insets. In the bottom left panel  we show the cluster size distribution $P(S)$ at the percolation threshold, the relative cumulative distribution $P(S>S^*)$ is displayed in the inset. In the right bottom panel we show the frequency-rank distribution $f(r)$ for the words in the model.}
 \end{figure}

In Fig.\ref{f5} we show that with an opportune choice of the parameters this model can generate a system with the desired properties.

%%%%%%%%%%%%%%%%%%%%%%%%%%%%%%%%%%%%%%%%%%%%%%%%%%%%%%%%%%%%%%%%%%%%%%%%%%Fig. 5

\section*{Discussion}
Nowadays understanding the topology of the SS and the dynamics of meaning is a fundamental issue in many  fields of knowledge and technology \cite{51}.
We can think about its value for understanding the dynamics of language and its evolution \cite{bs,33,60}, or its relevance in psycholinguistics and psychology \cite{35}.
Also, apart for being one of the main research topics in semiotics, linguistics and philosophy \cite{22}, it has recently been a hot topic in artificial intelligence and robotics \cite{34}.
Moreover there is an active effort in the information systems community to develop semantic-based web research tools \cite{db}.

The results of this research shed light upon the topology of the SS, represented as an  encyclopedic semantics.
The empirical research is unique in two fundamental aspects: the first one is the content analysis of a whole snapshot of Wikipedia, the second one is that this analysis is directional.

The empirical analysis reveals interesting properties of the SS. On one hand we can observe that the SS cluster size distribution is scale-free. This observation relates the SS to a wide range of scale-free phenomena \cite{49}. On the other hand we find a peculiar behaviour of the degree distribution. The latter observation relates the SS to a recently observed class of molecular biology phenomena, such as protein networks and more in general of genomic interaction networks \cite{48b,48c} and draws a bridge between phenomena whose underlying mechanism is a language, words for the SS and the DNA alphabet for the genomic networks.

 Models of content-based network are powerful tools and recently they have  attracted the attention of the scientific community especially for their relation to the so called \emph{hidden variable graphs} \cite{l2,l1}.
 The presented stochastic model is descriptive and  wants to individuate a few simple mechanisms that are able to reproduce some interesting statistical behaviours of the SS. In particular it is able to capture the statistical properties of Wikipedia at three levels of complexity. At  microscopic level it can reproduce the Zip's law for the word frequency distribution, at mesoscopic level it generates the scale-free distribution for the semantic cluster size  and finally at macroscopic level it can reproduce the exponents of the out and in-degree distribution.
   Interestingly enough we find again that the description of the SS at the model level resembles mechanisms of DNA evolution, where the process involved is of copy and mutation.

It is straightforward to relate the presented analysis with previous works on Wikipedia based on the  analysis  of the network of hyperlinks  connecting the Wikipedia different entries \cite{44}. In particular one of the questions arising from those works is if the Wikipedia hyperlink network has any relation to the underlying semantic network. Some attempts to answer  this question are exposed in \cite{51}, where a positive correlation is found between hyperlinked pages and their semantic content. In the light of this research we can say that the  topology of the SS is drastically different from the ones obtained by the hyperlinks analysis.
In the latter case the exponents of the degree distribution are smaller than -2 and linear preferential attachment is recovered \cite{44,45}, revealing a dynamics based on popularity. However this should not be a surprise since the network of hyperlinks is superimposed to the SS of the encyclopedia, so that it does not reflect the topology of the SS, but the structures locally imposed by the writers of the different entries.

We also notice that the topological properties of the SS are different from the ones obtained for dictionary semantics \cite{36,52}.
 The topological properties of dictionary networks, characterised by scale-free distribution with exponents smaller than -2, seem to be based again on a popularity mechanism and to reflect properties of language use more than the properties of the SS \cite{50}. In contrast  we find that the
architecture of SS is scale invariant, hierarchical, it has small-world properties, but it is not associated to a \emph{rich get richer} mechanism for the degree distribution.

In fact the SS structure is keen to be interpreted as an emerging property of a content-based network, where the Zipf's distribution  of the content words is a key feature for the resulting  topology.

% You may title this section "Methods" or "Models".
% "Models" is not a valid title for PLoS ONE authors. However, PLoS ONE
% authors may use "Analysis"
\section*{Materials and Methods}

\subsection*{A measure for the directional semantic flow}
Given two pages characterised by their lemmas frequency distributions $P$ and $Q$ and by their size $n_1$ and $n_2$, we define their distance $D(P\|Q)$ as:
\begin{equation} \label{1}
D(P\|Q)\equiv\frac{H(\pi_1P+\pi_2Q)-\pi_1H(P)-\pi_2H(Q)}{-\pi_1 \ln \pi_1-\pi_2 \ln \pi_2},
\end{equation}
where $H(P)=-\sum_i p_i\ln p_i$ is the Shannon entropy measured in
\emph{nats} and $\pi_i\equiv n_i/(n_1+n_2)$ are the weights.

If $D(P\|Q)\neq 0$,  the two pages have one or more words in common. Let's call $\overline P_J$  the frequency distribution of those common words for the first page, normalised to 1 and $\overline Q_J$  the frequency distribution of those common words for the second page, normalised to 1. Moreover let's call $\mu_P$ the fraction between the number of common words in the first page and the total number of words in the first page and $\mu_Q$ the fraction between the number of common words in the second page and the total number of words in the second page.
Then we can define a directionality index as:
\begin{equation}\label{2}
I(P\rightarrow Q)\equiv -\mathrm{sign}\left[\frac{H(\overline P_J)}{\mu_P} -
\frac{H(\overline Q_J)}{\mu_Q}\right].
\end{equation}

If $I(P\rightarrow Q)=1$ we can infer a direction of the semantic flow from the first page to the second one. Otherwise, if $I(P\rightarrow Q)=-1$, we can infer a semantic flow from the second page to the first one.

A detailed description of Eq.\ref{1} and Eq.\ref{2} is given in \cite{s2}

\subsection*{The model}
\emph{a}- When we generate a new page we first extract its size $l$ from a log-normal distribution, with first moment $\overline{l}$ and second moment $l_\sigma$. Then we fill the page with some new words from a potentially infinite vocabulary and some old words picked at random from the already written pages. To establish the balance between new and old words we assume a variation of the Heaps' law, frequently used in network theory \cite{40}, that states that the growth of the length $L$ of a written text is a quadratic function of the size of the vocabulary $t$,  $L(t)=\alpha t^2+t$. Then  we  assign to the page a random number $m$, so that $1<m<l$, representing the size of the invariant part of the page.

\emph{b}-When we mutate a page we keep unchanged the first $m-1$ words of the page, that is the page invariant part. For the last $l-m+1$ words of the page, each word is changed with probability $p$ and it is kept unchanged with probability $1-p$, where $p$ is a random number between 0 and 1. When we change a word we substitute it with an old or a new word considering the balance between vocabulary and text size as in point \emph{a}.

We start the model with a few pages of random words. Then at each time-step:

1- we generate a new page as explained at point \emph{a}.

2- We create $M$ new pages copying  $M$ old pages picked up randomly from the old pages and mutating them as explained at point \emph{b}.

The important parameters of the model are $M$ and $\alpha$. $M$ regulates the exponent of the cluster size distribution that goes, increasing the value of $M$, from $-3$ to $-2$. Moreover, increasing $M$,  more connected clusters form and this increases the exponent of the degree distribution from $-1.5$ to $-1$. The coefficient $\alpha$ regulates the amount of correlations between the different semantic clusters and this gives the possibility to tune the point of percolation of the system.

% Do NOT remove this, even if you are not including acknowledgments
\section*{Acknowledgments}
 Supported by Ministerio de Ciencia e Innovacion and Fondo Europeo de Desarrollo Regional through project FISICOS (FIS2007–60327). The authors are grateful to Konstantin Klemm, Marco Patriarca and Jos\'e Ramasco for the useful comments and discussions on the subject and to Chris Bate for his last proof-reading. The IGraph package was used for some of the analysis of this research \cite{ll1}.

%\section*{References}
% The bibtex filename
%\bibliography{template}

%\begin{figure}[!ht]
%\begin{center}
%%\includegraphics[width=4in]{figure_name.2.eps}
%\end{center}
%\caption{
%{\bf Bold the first sentence.}  Rest of figure 2  caption.  Caption
%should be left justified, as specified by the options to the caption
%package.
%}
%\label{Figure_label}
%\end{figure}

%\section*{Tables}
%\begin{table}[!ht]
%\caption{
%\bf{Table title}}
%\begin{tabular}{|c|c|c|}
%table information
%\end{tabular}
%\begin{flushleft}Table caption
%\end{flushleft}
%\label{tab:label}
% \end{table}

\end{document}